# Lessons learned after three years of SPIDER operation and the first MITICA integrated tests

D. Marcuzzi[a], V. Toigo[a], M. Boldrin[a], G. Chitarin[a,b], S. Dal Bello[a], L. Grando[a], A. Luchetta[a], R. Pasqualotto[a], M. Pavei[a], G. Serianni[a], L. Zanotto[a], R. Agnello[a,c], P. Agostinetti[a], M. Agostini[a], D. Aprile[a,d], M. Barbisan[a,c], M. Battistella[a], G. Berton[a], M. Bigi[a], M. Brombin[a], V. Candela[a], V. Candeloro[a], A. Canton[a], R. Casagrande[a], C. Cavallini[a], R. Cavazzana[a], L. Cordaro[a], N. Cruz[a,e], M. Dalla Palma[a], M. Dan[a], A. De Lorenzi[a], R. Delogu[a], M. De Muri[a], M. De Nardi[a], S. Denizeau[a], M. Fadone[a], F. Fellin[a], A. Ferro[a], E. Gaio[a], C. Gasparrini[a], F. Gnesotto[a], P. Jain[a], A. La Rosa[a], D. Lopez-Bruna[a,g], R. Lorenzini[a], A. Maistrello[a], G. Manduchi[a], S. Manfrin[a], N. Marconato[a], I. Mario[a,t], G. Martini[a], R. Milazzo[a], T. Patton[a], S. Peruzzo[a], N. Pilan[a], A. Pimazzoni[a,d], C. Poggi[a], N. Pomaro[a], B. Pouradier-Duteil[a,c], M. Recchia[a], A. Rigoni-Garola[a], D. Rizzetto[a], A. Rizzolo[a], F. Santoro[a], E. Sartori[a,b], B. Segalini[a], A. Shepherd[a,h], M. Siragusa[a], P. Sonato[a], A. Sottocornola[a], E. Spada[a], S. Spagnolo[a], M. Spolaore[a], C. Taliercio[a], P. Tinti[a], P. Tomsič[i], L. Trevisan[a], M. Ugoletti[a], M. Valente[a], M. Valisa[a], F. Veronese[a], M. Vignando[a], P. Zaccaria[a], R. Zagorski[a,l], B. Zaniol[a], M. Zaupa[a], M. Zuin[a], M. Cavenago[d], D. Boilson[m], C. Rotti[m], H. Decamps[m], F. Geli[m], A. Sharma[m], P. Veltri[m], J. Zacks[m], M. Simon[n], F. Paolucci[n], A. Garbuglia[n], D. Gutierrez[n], A. Masiello[n], G. Mico[n], C.Labate[n], P.Readman[n], E.Bragulat[n], L.Bailly-Maitre[n], G.Gomez[n], G.Kouzmenko[n], F.Albajar[n], M. Kashiwagi[o], H. Tobari[o], A. Kojima[o], M. Murayama[o], S. Hatakeyama[o], E. Oshita[o], T. Maejima[o], N. Shibata[o], Y. Yamashita[o], K. Watanabe[o], N.P. Singh[p], M.J. Singh[p], H. Dhola[p], U. Fantz[q], B. Heinemann[q], C. Wimmer[q], D. Wünderlich[q], K. Tsumori[s], G. Croci[t], G. Gorini[t], A. Muraro[u], M. Rebai[u], M. Tardocchi[u], L. Giacomelli[u], D. Rigamonti[u], F. Taccogna[u], D. Bruno[u], M. Rutigliano[u], S. Longo[u], S.Deambrosis[v], E.Miorin[v], F.Montagner[v], A. Tonti[w], F. Panin[z]

[a]Consorzio RFX, Corso Stati Uniti 4, I–35127 Padova, Italy
[b]Università degli Studi di Padova, Dept. of Management and Engineering, Strad. S. Nicola 3, 36100 Vicenza, Italy
[c]Ecole Polytechnique Fédérale de Lausanne (EPFL) - Swiss Plasma Center (SPC), 1015 Lausanne, Switzerland
[d]INFN-Laboratori Nazionali di Legnaro (LNL), v.le dell'Università 2, I-35020 Legnaro PD, Italy
[e]Instituto de Plasmas e Fusão Nuclear, Instituto Superior Técnico, Universidade de Lisboa, 1049-001 Lisboa, Portugal
[f]Institute for Nuclear Research and Nuclear Energy, Bulgarian Academy of Sciences, 72 Tsarigradsko Chaussee Boulevard, Sofia, 1784, Bulgaria
[g]Laboratorio Nacional de Fusión, CIEMAT, Madrid, Spain
[h]CCFE, Culham Science Centre, Abingdon, Oxon, OX14 3DB, UK
[i]University of Ljubljana, Faculty of Mechanical Engineering
[l]National Centre for Nuclear Research (NCBJ), PL-05-400 Otwock, Poland
[m]ITER Organization, Route de Vinon-sur-Verdon, CS 90 046, F-13067 St. Paul-lez-Durance, France
[n]Fusion for Energy,C/o Josep Pla 2, E-08019 Barcelona, Spain
[o]National Institutes for Quantum and Radiological Science and Technology (QST), 801-1 Mukoyama, Naka, Ibaraki-ken 311-0193, Japan
[p]ITER-India, Institute for Plasma Research, Nr. Indira Bridge, Bhat Village, Gandhinagar, Gujarat 382428, India
[q]IPP, Max-Planck-Institut für Plasmaphysik, Boltzmannstraße 2, D-85748, Garching bei München, Germany
[r]LAPLACE, Université de Toulouse, CNRS, Toulouse, France
[s]National Institute for Fusion Science, 322-6 Oroshi, Toki, Gifu 509-5292, Japan
[t]Dipartimento di Fisica "G. Occhialini", Università di Milano-Bicocca, Milano, Italy
[u]ISTP-CNR, Institute for Plasma Science and Technology, Via Roberto Cozzi 53 - 20125 Milano (MI)
[v]National Research Council of Italy - CNR, Institute of Condensed Matter Chemistry and Technologies for Energy – ICMATE, Corso Stati Uniti, 4, 35127 Padova, Italy
[w]INAIL-DIT, Via Ferruzzi, 40 - 00143 Roma
[z]INAIL-UOT Padova, Via Nancy, 2 - 35131 Padova

ITER envisages the use of two heating neutral beam injectors plus an optional one as part of the auxiliary heating and current drive system, to reach the desired performances during its various phases of operation. The 16.5 MW expected neutral beam power per injector is several notches higher than worldwide existing facilities.
In order to enable such development, a Neutral Beam Test Facility (NBTF) was established at Consorzio RFX, exploiting the synergy of two test beds, called SPIDER and MITICA. SPIDER is dedicated developing and characterizing large efficient

---

*author's email: diego.marcuzzi@igi.cnr.it*

negative ion sources at relevant parameters in ITER-like conditions: source and accelerator located in the same vacuum where the beam propagates, immunity to electromagnetic interferences of multiple radio-frequency (RF) antennas, avoidance of RF-induced discharges on the outside of the source. Three years of experiments on SPIDER have addressed to the necessary design modifications to enable full performances. The source is presently under a long shut-down phase to incorporate learnings from the experimental campaign, in particular events/issues occurred during operation, which led to the identification of improvement opportunities/necessities (e.g. RF discharges, local burns, water leaks, other damages, configuration/design upgrades to maximize chances/margin to quest target parameters).

Parallelly, developments on MITICA, the full-scale prototype of the ITER NBI featuring a 1 MV accelerator and ion neutralization, are underway including manufacturing of the beam source, accelerator and the beam line components, while power supplies and auxiliary plants, already installed, are under final testing and commissioning.

Integration, commissioning and tests of the 1MV power supplies are essential for this first-of-kind system, unparalleled both in research and industry field. 1.2 MV dc insulating tests of high voltage components were successfully completed. The integrated test to confirm 1MV output by combining invertor systems, DC generators and transmission lines extracted errors/accidents in some components. To realize a concrete system for ITER, said events have been addressed and solutions for the repair and the improvement of the system were developed.

Hence, NBTF is emerging as a necessary facility, due to the large gap with existing injectors, effectively dedicated to identify issues and find solutions to enable successful ITER NBI operations in a time bound fashion. The lessons learned during the implementation on NBTF and future perspectives are here discussed.

## 1. Introduction

The ITER Neutral Beam Test Facility (NBTF) is targeted at the development and testing up to nominal performances of the full-scale prototype of the ITER Heating Neutral Beam Injector (HNB) [1,2,3].

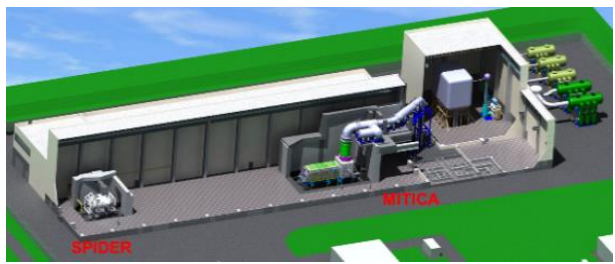

Fig. 1 – The Neutral Beam Test Facility, with SPIDER and MITICA experiments

The NBTF is established in Padova (Fig.1), integrated in an international framework of responsibility and collaboration among Consorzio RFX, ITER Organization (IO), Fusion for Energy and JADA/QST with the involvement of some European research institutes; it is strictly liaised with the Indian Test Facility (INTF) in charge of the ITER Diagnostic Neutral Beam (DNB) [4].

Two test beds are foreseen on NBTF site: SPIDER features the first prototype of ITER ion source, with strong similarities with DNB configuration, while MITICA is the 1:1 scale prototype of ITER HNB.

In 2021, after three years of operation, SPIDER entered a long shut down phase with the aim of solving issues arising and detected during the first operational phase. MITICA is in the final construction phase of the injector, with all the auxiliary systems installed and with the final integrated tests of the overall power supply underway.

Efforts are being paid in synergy with other experiments worldwide focusing on negative ion based neutral beams, in particular in Japan and India, in relation to their direct involvement in ITER and in the NBTF, and also in Europe, supported by EUROfusion, for the challenging RF source development.

## 2. SPIDER

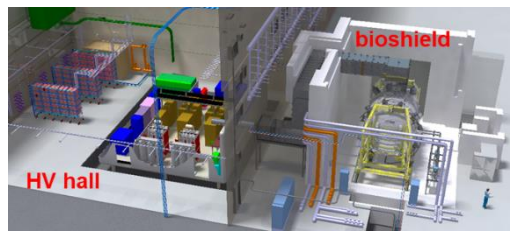

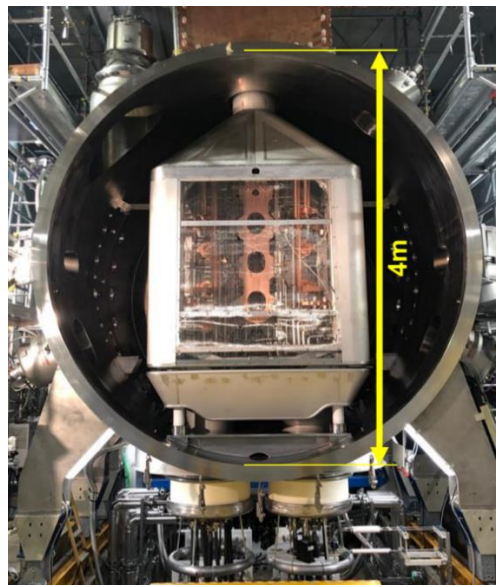

Fig. 2 – SPIDER experiment

SPIDER experiment (Fig.2), the third leg in EUROfusion step ladder approach towards ITER source for NB injectors, together with BUG and ELISE at IPP-Garching [5], represents a novelty in the negative ion

_________________________________________________
*author's email: diego.marcuzzi@igi.cnr.it*

sources for many aspects related not only to the dimensions. Among the most impacting factors, it is necessary to highlight that, unlike any other existing NBIs, the entire beam source of SPIDER (and MITICA) lies inside the vacuum vessel, so that it is surrounded by the same background gas at low pressure, in which the beam propagates. This configuration is required for HV holding and specifically in ITER for tritium containment.

The SPIDER testbed is operational since 2018. A number of occurrences characterized SPIDER operation, leading to the identification of early issues that in part were solved or worked around on the run, and other have triggered requirements for component re-design or modification and opportunities for improvements.

### 2.1. Main recent experimental results

field and biasing voltage for source and bias plate, exploiting in full the capabilities of the large set of diagnostics put in place.

Then, the first campaign featuring Cs injection was carried out: the production of negative ions in the proximity of the plasma grid benefitted from the evaporation of caesium into the source (fig.4), as expected, and the beamlet divergence decreased with respect to the operation without caesium.

It was recognized that the negative ion current density and the current of co-extracted electrons, in the same conditions of RF power, gas pressure and bias voltages, depend on the ratio between the caesium evaporation rate and the plasma on-off duty cycle; the time evolution of H-density revealed the importance of the magnetic filter field in preserving the caesium layer on the plasma grid

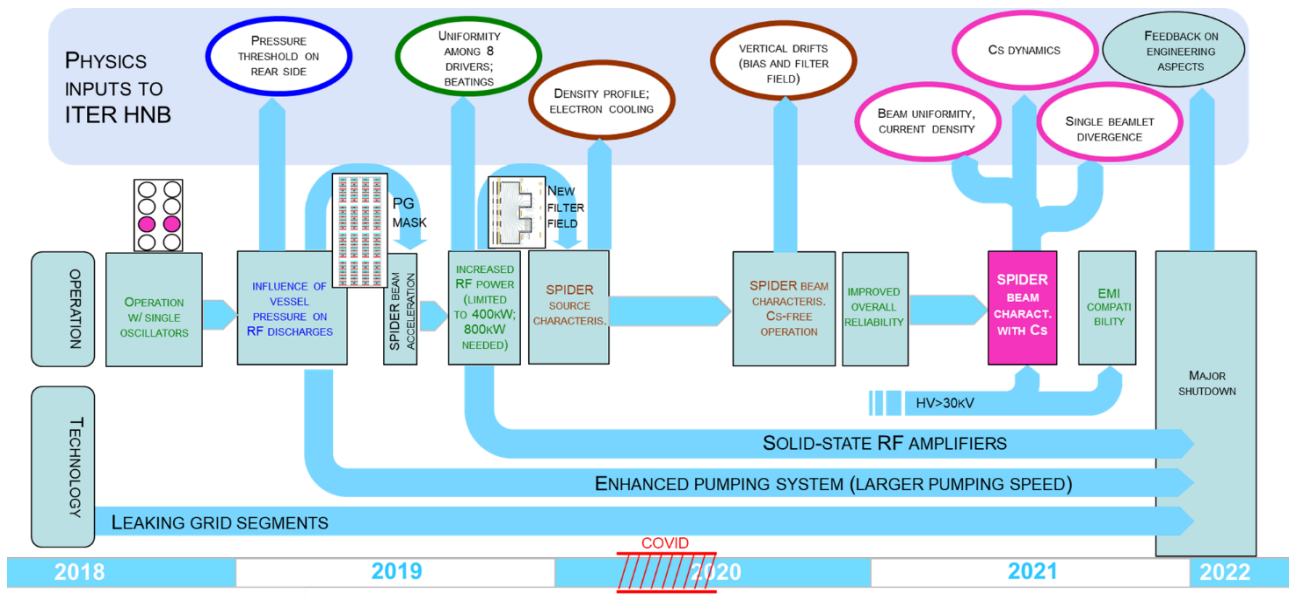

Fig. 3 – Summary of SPIDER experimental campaign

In these first three years, SPIDER operations generated a wealth of experimental information, which provided insight into the source performance and evidenced operational issues that must be resolved in view of extensive beam operation and particularly of MITICA. About 250 experimental days were carried out overall, the final experimental Cs campaign consisting of 19 days (2 of which in D); overall around 4000 pulses were executed for an order of 60 h pulse on time.

Fig.3 summarizes SPIDER experimental campaign so far, highlighting main events, outputs and implications for the shutdown.

During the last campaign without Cs injection, the improved maturity of the system allowed to increase the performances, in particular the acceleration voltage; a mask on Plasma Grid (PG), allowing only few beamlets to be extracted, was temporarily implemented in order to prevent RF discharges, occurred in the beginning of the experimental phase [6].

It was possible to perform several scans to characterize the beam optics in volume up to 50kV, both in hydrogen and deuterium, with different settings of filter from the action of plasma.

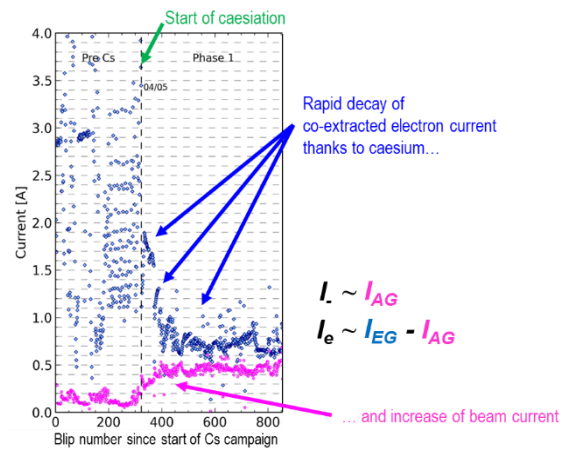

Fig. 4 – Immediate effect of first cesiation on electron and ion current. In x axis the pulse number is reported, while the ion and electron current are reported in y axis, being IEG and IAG the currents driven by extraction and acceleration grid power supplies, respectively.

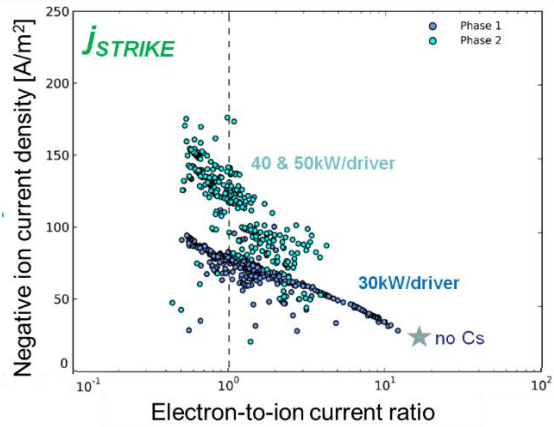

Fig. 5 – Electron-ion ratio and negative ion current density in the first cesium campaign (in hydrogen), compared with no-Cs result (SPIDER requirement is 0.5 for e/H$^-$ ratio and 355 A/m$^2$ in H); the negative ion current density is measured by the instrumented calorimeter STRIKE [7].

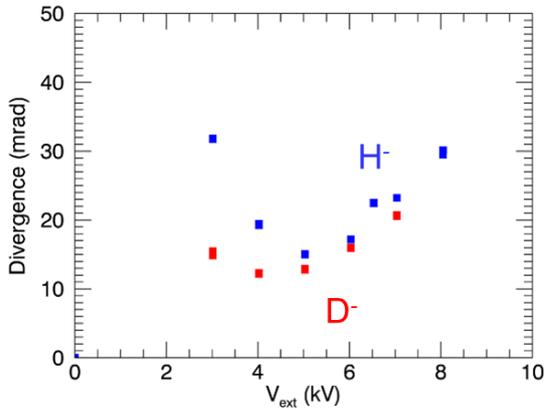

Fig. 6 – Beamlet divergence in the first cesium campaign, measured by Beam Emission Spectroscopy [8] with typical accuracy < 10% (ITER requirement is 7 mrad)

While for the ratio of co-extracted electrons to ions obtained results already match the target range (fig.5), it has been recognized that ion current density, beam uniformity and divergence (fig.6) require further efforts in the future campaign to reach the objectives, that will require to exploit in full the RF generators [9].

The next sections focus mainly on SPIDER configuration refurbishments, improvements and next actions identified from the practical findings and results, more detailing engineering implications, rather than the physics background and the expected functional perspectives for HNBs, for which some preliminary hints are presented in [10].

### 2.2. Shutdown activities

The SPIDER shutdown started at the end of 2021, extracting the source from the vessel, then carrying out a full disassembly and inspection of the status of individual source parts.

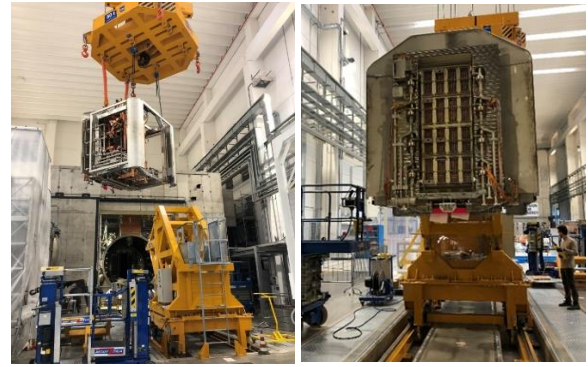

Fig. 7 – Removal of SPIDER source

During inspection after disassembly many interesting details have been noted, bringing further inputs for the shutdown activities (fig.8).

Throughout the disassembly process, initial analyses of the machine and components were continuously performed and are still being done: chemical analyses of surface depositions; magnetic field measurements of the permanent magnets embedded in the drivers, source walls and grid segments; He leak tests at hydraulic connections; grid relative positions by optical metrology; plasma facing surface state scanning and analysis.

Notable degradation of some parts was recorded and is under careful assessment: general surface condition of the source main chamber (effects from interaction with plasma, residues after a water leak, localized Mo layer wear or detachment), erosion effects due to electrical discharges on the rear side of the source, damages to some alumina parts (bushings, washers, insulators) on driver back cover fixation and grid segments interfaces, damages on the bottom four drivers [11].

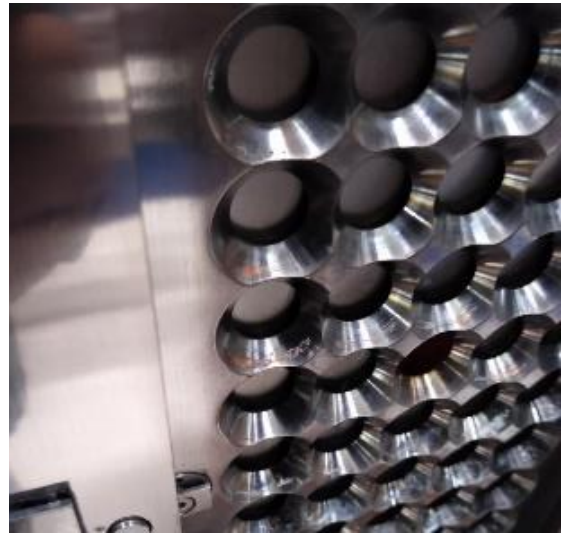

Fig. 8 – Example of disassembly and inspection of source parts (to be noted, copper peeking out from under molybdenum coating on PG)

Nevertheless, the main goal of the shutdown focusses on the planned substantial changes:

- already identified modification of the beam source with RF drivers and layout enhancements, extraction grid changes and other minor upgrades; the replacement of the damaged components discovered during disassembly activities have been included in the planning;

- replacement of RF oscillators with solid state amplifiers;
- upgrading of the pumping system to be able to operate the source up to 0.45 Pa of pressure with a Pvessel ≤ 40 mPa in order to avoid the risk of discharges induced by the RF electric field, with no grid masking, hence all beamlets active;

During the shutdown other activities will also be carried out, including maintenance and implementation of minor improvements to power supply and other auxiliary systems, updating of diagnostic systems [12,13,14,15].

At the end of the shutdown SPIDER is expected to be ready to proceed towards full performances.

### 2.3. Source modifications

The list of the main issues occurred on source components during operations and the corresponding remedy foreseen during the shutdown is listed in Fig.9.

| component | main issues | remedy |
|---|---|---|
| SPIDER RF Drivers enhancement | local high Efield large loop areas | revision of design (coil turns, connections, part shape & |
| SPIDER GG replacing segments procurement | water leak | new segment |
| SPIDER EG modifications | measurement of single segment current | insulated segments tapered apertures |
| Recovery of PG filter busbar electrical insulation | damage to the plasma Al2O3 plasma spray | modified design with solid insulating shims |
| Remedial actions for melted rings around drivers holes | melted fixing ring due to eddy currents | replacement with discrete brackets |
| RF Drivers FSLWs procurement | 4 damaged FSLWs | Partial re-manufacturing |
| Molibdenum coating refurbishment | Some worn/damaged plasma facing surfaces | De-coating and re-coating of parts |

Fig. 9 – Summary of the main events and limits identified so far in SPIDER source operation. Further investigations on possible limits form the functional/physics point of view are still ongoing, hence not dealt with in detail in this paper.

The activities for the refurbishments and the modifications have been planned and organized each as in single project, in order to better control the progress status and to follow-up the procurements:

- RF Driver enhancement: in addition to the improvement to the RF circuit layout and connections onboard the beam source, other modifications were designed and implemented, like electrostatic screens for shielding triple-points on capacitors and ceramic breaks, shown in Fig.10, and a new less cumbersome bias busbar on the back of RF source [16].

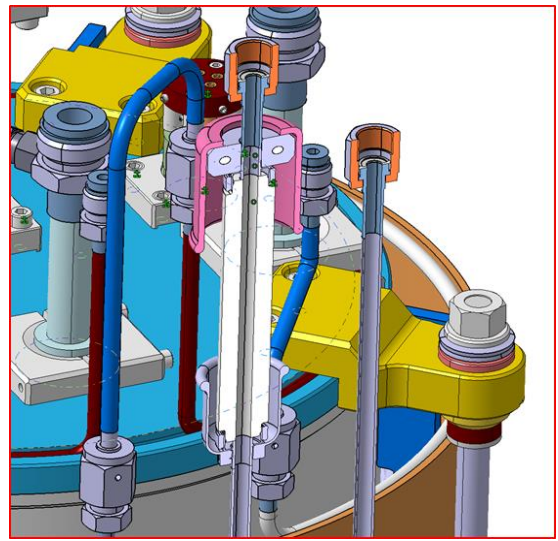

Fig. 10 – Example of modification: addition of a local electrostatic shield (in pink)

- RF circuit layout improvement to reduce the mutual coupling between different driver pairs, which was found responsible for the modulation of the plasma light when two or more driver pairs were supplied [17]

- Extraction Grid (EG) modifications: EG segments will be modified to be electrically insulated among themselves, in order to discriminate the electrical current collected by each segment from extracted beam, hence to identify excessive non-uniformities potentially dangerous for the segment integrity; moreover, the EG apertures will be tapered in order to increase the margin preventing the beamlets from colliding with the aperture' rims. See Fig. 11.

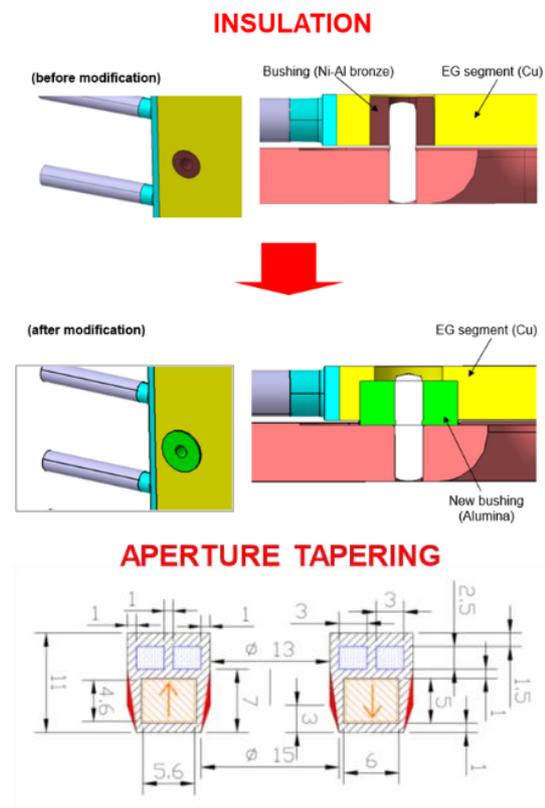

Fig. 11 – Modifications to EG segments

- Plasma Grid (PG) frame cooling enhancement: modification to the already existing cooling system of the PG support flange in order to improve its efficiency so to allow to operate safely at steady state with PG thermalized at 150°C.

- Recovery of PG filter busbar electrical insulation: modification to the PG filter circuit busbars embedded in the rear driver plate (RDP) and electrically insulated by a layer of alumina coating. The new design foresees bare busbars (no more alumina coated) with peek elements guaranteeing the distance from the RDP surfaces and so the electrical insulation.

- Procurement of a new grounded grid segment to replace the leaky one

Most of SPIDER source modifications are MITICA/HNB relevant: some of them were already known and have already been integrated in MITICA during the ongoing procurement, while for the others an experimental validation is expected before implementing the transfer.

### 2.4. RF power supply generators

The original configuration of RF systems includes 4 generators based on free-running tetrode oscillators in the push-pull self-excited Colpitts configuration [17,18] supplying 4 driver pairs on board the ion source, used to ionize the gas and excite plasma. Four Normal L-type matching networks are mounted on the backside of the ion source, to tune the driver pair impedance to that of the RF generators [19]. With this matching network the load impedance (i.e. the RF transmission line plus the ion source impedance) is equal to the output impedance of the oscillator, the so-called matching condition, only on a narrow frequency range. The achievement of the matching condition is of paramount importance to deliver power to the load without stressing the entire RF system, and it is achieved by varying the operating frequency of the generators.

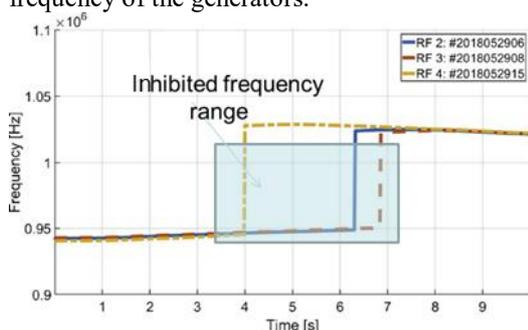

Fig. 12 – Frequency flips with RF oscillators, impeding to reach the best matching conditions with plasma impedance

Since the beginning of the experiments in SPIDER, however, the operation of the RF oscillators has been hindered by some important limitations. First of all, the appearance of the so called "frequency flips" [20], which are frequency instabilities intrinsic in the operation of the oscillator on a resonant load (Fig.12) preventing the achievement of the matching conditions [21]. Even though the mechanism of their onset has been understood and modelled in SPIDER, frequency flips are unavoidable and results in strong limitations of the performance in terms of power delivered to the source [22].

In addition to the frequency flip issue, the presence of relatively high voltage in the oscillator requires conditioning procedures for high voltage vacuum capacitors to avoid internal discharge during operation; the conditioning may be a long process, reducing the availability of the system. High voltage means also that maintenance is often required for the most critical components.

The above-mentioned evidences led to the decision to replace the RF generators in SPIDER and MITICA and change the current ITER baseline from self-excited free running tetrode oscillator technology to solid-state amplifier technology [17,23,24]. The solid-state amplifier technology is believed to be strongly attractive for ion source operation, since no frequency flip phenomena are expected and there are no tetrodes requiring high voltage. In fact, at IPP (Garching) solid-state generators have already been tested with positive feedback [25,26].

### 2.5. Vacuum system enhancement

Ever since the initial phase of the experimental campaign, the need to enhance SPIDER vacuum pumping capability became evident, as a consequence of different contributing factors: the measured higher conductance of the beam source with respect to the design value, the presence of RF-induced breakdowns on the rear side of the beam source at high pressure in the vessel (greater than 40mPa) and the increased max working pressure in the source ($\geq 0.4$Pa), as it would facilitate the negative ion production.

All these aspects, which are linked, will require the installation of an additional pumping system to be operated in parallel to the existing one.

A preliminary assessment allowed to compare the possible solutions to integrate additional pumping capacity, among commercial cryopumps, customized cryopanels or Non-Evaporable Getter (NEG) pumps.

The chosen solution consists of the NEG pump system (see fig.13), based on the Non-Evaporable Getter cartridges of ZAO® alloy, developed by SAES Getters company [27,28], basically because a suitable system of commercial cryopumps would be too cumbersome, while customized cryopanels would require an expensive additional dedicated cryogenic system and leading to a very complex integration.

The cartridge is NEG elementary unit, composed by disks organized in stacks. The cartridges are arranged on panels with a regular pattern, to be installed inside the SPIDER vessel, providing a pumping speed S=215-240 $m^3$/s with the baseline configuration of 384 cartridges, potentially extendable to S=260-280 $m^3$/s with 512 cartridges; capacity for hydrogen/deuterium is 240000/170000 Pa*$m^3$ with a regeneration time lower than 12 h.

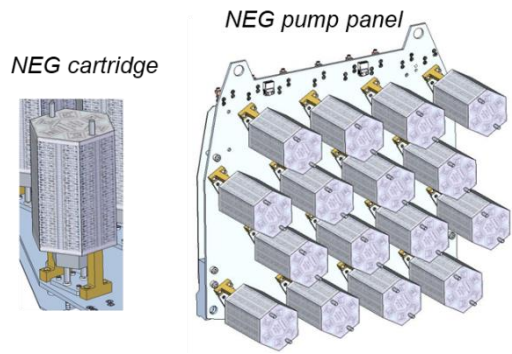

Fig. 13 – NEG pump basic elements, cartridge and panel

The system is completed with the power supply necessary to heat the NEG elements for activation and regeneration and the local control unit needed to operate the pump.

An additional cylindrical sector is needed to host the NEG system by extending the existing SPIDER vessel; moreover, thermal electrostatic shields shall be included with the double scope to protect the NEG pump from electrical discharges occurring on the beam source, and to protect all the components surrounding the NEG pump from thermal radiation during regeneration, requiring high temperature for the cartridges. All these new elements will be actively cooled by a dedicated water circuit, as shown in Fig.14 [29].

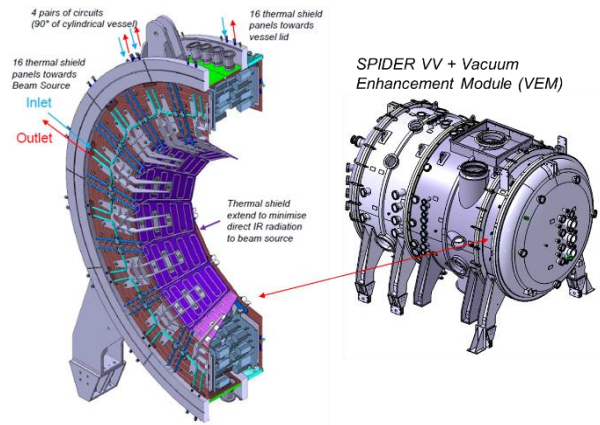

Fig. 14 – Additional vessel module with NEG panels and actively cooled thermal shields

MITICA is not impacted by this issue/improvement, as the vacuum is provided by the large cryopanels.

### 2.6. Other sub-systems

Indeed, there were also some sub-systems that proved to be quite effective already in the short term after their integration, hereafter a pair of examples.

The Cs oven system is necessary to inject Cs inside the source to enhance negative ion production. It is quite a multidisciplinary and spread system, comprising cubicles inside the HVD for the power supplies, the diagnostic racks and the PLC periphery (I/O modules) that are connected to the ovens mounted on board of the source in vacuum by means of the transmission line, at the beam source potential. Another cubicle in the I&C room hosts the local control panel, the PLC CPU and the connections to CODAS and Interlock, as shown in Fig. 15. Before the operation on SPIDER a Cs oven prototype and the 3 definitive ovens have been tested and characterized on the CATS (CAesium Test Stand) facility at NBTF [30,31]. After having successfully completed the installation and commissioning, demonstrating compatibility and resilience to EMI and noise, each oven was ready to operate, filled with 10g of metallic Cs. Cs evaporation has been successfully performed at different flow rates (6 – 20 mg/h per oven) throughout the whole campaign and its effect on the co-extracted electrons and negative ion generation has been shown in [6,11]. Cs flow is constantly measured by means of a Surface Ionization Detector (SID) placed directly on the injection nozzle. Finally, during the oven inspection at the end of the campaign, contaminated Cs was found in the reservoirs. This probably was due to the massive water leak occurred at last causing water vapor to react with the remaining Cs and impurities, so compromising the leak tightness of the valve of the oven. Therefore, the total quantity of evaporated Cs during the experimental campaign could not be inferred by direct weighting of residual Cs in the reservoir, but it can only be estimated by the SID data. SID measurements are sufficiently accurate as shown in [11] and, specifically, they show that a total of 2.4g of Cs has been evaporated by the 3 ovens. This first Cs campaign does not allow an extrapolation reliable enough for a robust assessment of ITER HNB consumption of Cs; nevertheless, the numbers obtained so far would confirm the correct design for the overall ovens capacity (3*40=120g).

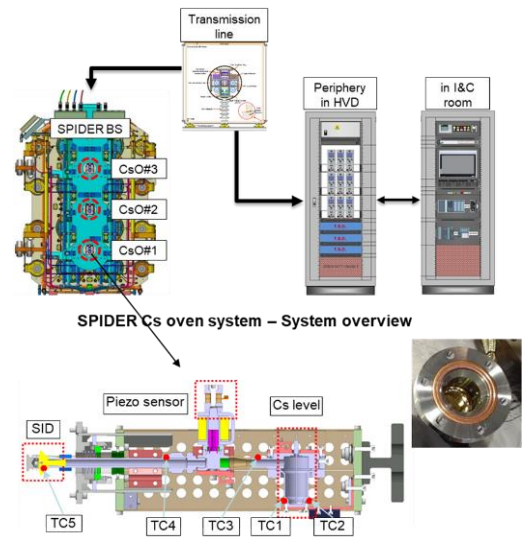

Fig. 15 – SPIDER Cs oven

The suite of diagnostics with complementary capabilities dedicated to the SPIDER source and beam proved essential to characterize the performance and identify the still open issues. The beam diagnostics used so far are illustrated in Fig.16 [32,33,34,35,36,37,38]: STRIKE uncooled calorimeter, with high spatial resolution, measures the single beamlet intensity profile, visible tomography and beam emission spectroscopy derive the beamlets intensity and divergence respectively without intercepting the beam; current monitors of individual beamlets measure also the high frequency components of the beam intensity and the Allison scanner provide the full emittance of specific beamlets. The last two systems were not part of the initial set and have been developed and installed during the experimentation phase, to validate and extend the capabilities of the other systems. Other diagnostics will be introduced in the current shutdown, mainly to better investigate the non-uniformity in the plasma source both locally inside a beamlet group and globally over the source extension [39,40,41,42]. Among them, insertable probes are foreseen to measure the plasma parameters along a driver axis up to the grids and parallel to the plasma grid; a retarded field energy analyzer will measure the positive ions energy distribution at the plasma grid; Fiber Bragg Grating sensors will replace thermocouples where they were strongly affected by electromagnetic noise.

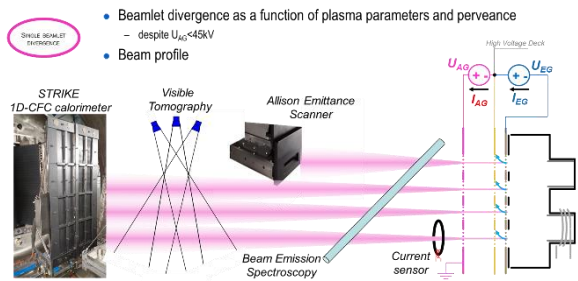

Fig. 16 – Single beamlet characterization

## 3. MITICA

MITICA, the one-to-one prototype for ITER HNB, is gradually approaching the start of operation.

Prior to that, three main areas of work are being dealt with in order to complete the whole experiment assembly, testing and commissioning:

- Last components/plants still under completion [43]
- Power supply integrated final tests and specific commissioning
- Preparation for a dedicated campaign of HV holding tests in vacuum

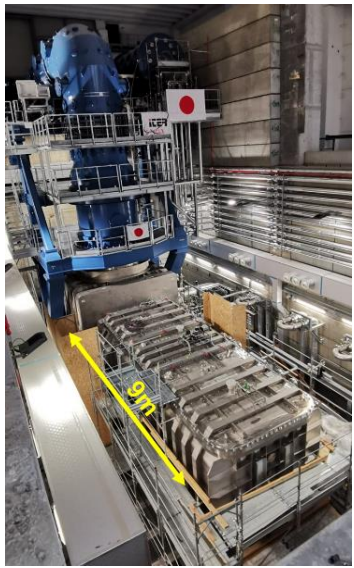

Fig. 17 – Current status of MITICA injector vessel

### 3.1. Ongoing procurements

While power supply system installation on site has been completed some years ago to allow the necessary long testing phase, other components and auxiliaries are still under procurement or installation activities:

- After vessel installation, the procurement is still ongoing for in-vessel mechanical components, such as beam source [44], beam line components and cryopumps;
- Auxiliary plants procurement under completion (F4E procurement), like cryogenic plant
- Last one to one "single-plant"/CODAS integrated commissioning are ongoing or expected soon

### 3.2. Integrated power supply tests

MITICA power supply is a very complex system, beyond the limit of conventional industrial technologies [45]. It is the first prototype developed in the world at 1MVdc with such power rating. One major purpose for MITICA is to develop, test and fine-tune this electrical system.

Designed by the European and Japanese teams under IO coordination, and based on international Codes and Standards and previous experience at JT60 (Naka – JA) plus additional specific R&D, the high voltage components for power supplies were procured by JADA and F4E. Each team supplied its share, including installation onsite and standalone site acceptance tests.

Nevertheless, at the end single components are strongly integrated during the operation and this required a deep coordination for the interfaces and an integrated commissioning process, very complex in technical terms and also responsibility-wise.

After successful completion of low power insulation and functional tests, in 2021 failures occurred during the 1MV high power integrated tests due to voltage breakdowns that damaged two components of the power supply system [46,47,48,49]:

- one branch of the rectifier diode bridge of the stage at 800kV-1MV (1 MVdc is the result of the series connection of five 200 kVdc stages)
- the HV bushing of the 1 MV insulating transformer which fed the ion source.

Before the faults, the integration tests of MITICA power supply have passed the level of 700 kVdc for 1000s.

Inspections and numerical analyses using fast transient models were carried out to determine the root cause of these unexpected events related to this very high voltage level and very difficult to evaluate during the design phase for the lack of experience: the dynamics of failures due to the voltage breakdown were explained, and solutions to increase the reliability and availability of the system have been identified. They will be implemented in MITICA during the restoring phase. Very useful information for improving the design of the power supplies of ITER has been derived also, triggering a review of the existing project to include additional protections.

However clear evidence of the exact location of the BDs was not found yet.

Hence, currently the overall activity, undertaken by the QST, NBTF team and ITER, is proceeding on parallel lines:

- execution of low power insulation test campaigns to find the exact position of the voltage breakdown and increase the tightness of the system's insulation, also exploiting additional diagnostics and monitoring system, developed with the support of experts from a high voltage laboratory;
- development of the restore and improvement plans for the diode bridges and the 1MV insulating transformer with the introduction of additional protection systems.

As a lesson learned, the explanation of the two faults was obtained by means of fast transient models which take into account the parasitic parameters C and L prevailing in the case of very rapid dynamics (of the order of MHz) and which depend on the real geometry of the components which could not be known in advance as it depends on the project developed by the supplier and which is part of its know-how. It cannot be forgotten that this is a 1MV dc system unique in the world for which there was no previous operating experience.

The reference scheme was based on the past experience of the J60SA, where a 500kV negative NBI, the most energetic NBI in the world, operated for several years without any power system problems. Moreover, the individual components of the electrical power system have been designed and tested by applying international standards with appropriate safety margins similar to those adopted in the 500kV system [48].

### 3.3. Preparations for HV tests in vacuum

A key point in the development of the ITER injector is the 1 MVdc voltage holding in the vessel.

HV test campaigns on MITICA experiment is foreseen before the installation of the in-vessel components, with the aim at gaining insights to one of the main topics to be investigated by the MITICA testbed.

The main objectives of these tests are to:

- guarantee the voltage holding of MITICA up to -1 MV in vacuum and low-pressure gas, before the installation of the Beam Source;

- establish and validate Voltage Holding scaling laws for large gaps and multiple electrodes (required for effective design optimization).

A project was launched to cope with the design, procurement and installation of an electrostatic shield (see fig. 18), reproducing the beam source surface, in time for tests in advance with respect to the delivery of the actual source [50,51].

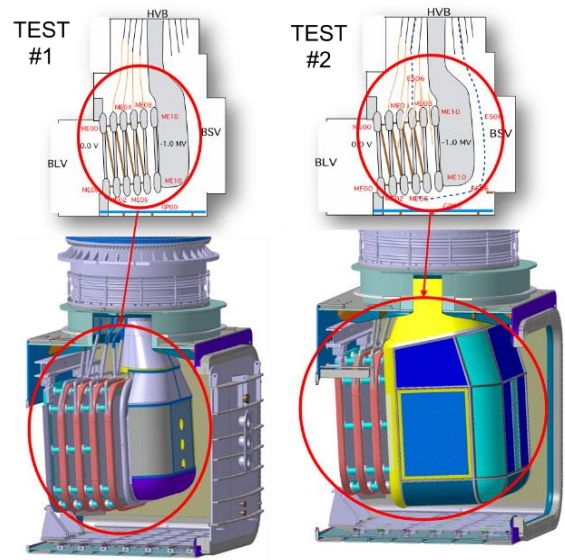

Fig. 18 – Configurations for HV tests in vacuum (source mockup evidenced)

Experiments carried out at QST showed the likely need of at least one intermediate shield at intermediate voltage in order to enhance the voltage holding between ion source at -1 MV and the grounded vessel. An intermediate electrostatic shield that will be installed on the -600 kV stage has been designed, in order to be tested on the mockup and also reused on the actual beam source, if confirmed necessary. The design considers the electrostatic purposes and the conflicting need to allow evacuation of the volume around the outer rear side of the source, resulting in a double-skin configuration with staggered patterns of holes (see Fig.19).

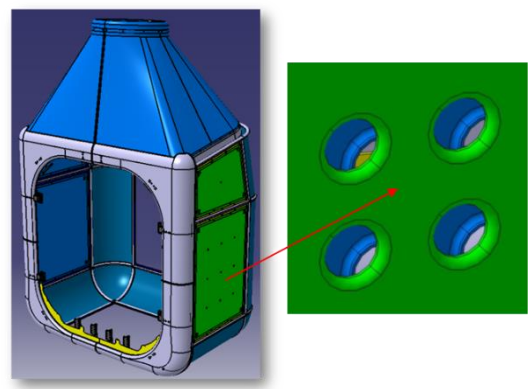

Fig. 19 – Design of the intermediate electrostatic shield

In addition, ancillary elements have been identified and added to perform the voltage holding tests: vacuum pumping system monitoring cameras and integration of electrical measurements and dedicated control system.

### 4. Conclusions

SPIDER is entering the first long shutdown after the first three years of experiments. Interesting results have been obtained at reduced performances, which have also identified necessary modifications, mainly focused on source upgrade, RF generators replacement and enhancement for the vacuum pumping system, in order to quest the target performances during the next operational phase.

MITICA is proceeding towards the end of preparatory activities prior to the start of the first experimental campaign. Some voltage breakdowns occurred during final integrated tests of the power supply system. Thanks to the new models developed ad-hoc, remedial actions and additional protection systems are under implementation in MITICA as well as in the design of ITER NB injectors.

Such voltage breakdown events have showed the lack of present experience and data on the 1 MVdc supply systems and MITICA is emerging necessary to cover this gap in advance of the ITER operation.

In parallel to the completion of the procurements for MITICA in-vessel components, preparations for the execution of HV tests in vacuum are ongoing to investigate the voltage holding capability prior to the first experimental campaign.

Overall, the most positive aspect of these experiences on both SPIDER and MITICA is that the facility is fully fulfilling its role, highlighting the critical issues of the project and allowing to intervene before the system is implemented on ITER.

The already ongoing tight collaboration between NBTF team and IO allows to take into account the solutions identified for SPIDER and MITICA in view of implementation on ITERR NB sustem.

## Acknowledgement and Disclaimer

This work has been carried out within the framework of the ITER-RFX Neutral Beam Testing Facility (NBTF) Agreement and has received funding from the ITER Organization.

This work has also been carried out within the framework of the EUROfusion Consortium, it has been funded by the European Union via the Euratom Research and Training Programme (Grant Agreement No 101052200 — EUROfusion).

Views and opinions expressed are however those of the author(s) only and do not necessarily reflect those of the European Union or the European Commission. Neither the European Union nor the European Commission can be held responsible for them. Also they do not necessarily reflect those of the ITER Organization.